\RequirePackage[hyphens]{url}
\documentclass[sigplan,screen]{acmart}
\renewcommand\footnotetextcopyrightpermission[1]{} 
\def\BibTeX{{\rm B\kern-.05em{\sc i\kern-.025em b}\kern-.08emT\kern-.1667em\lower.7ex\hbox{E}\kern-.125emX}}
\usepackage{dblfloatfix}
\usepackage{subfigure}
\usepackage{verbatim}
\PassOptionsToPackage{hyphens}{url}
\usepackage[utf8]{inputenc}

\begin{document}

%
\title[Adaptive Heterogeneous Scheduler]{An Adaptive Performance-oriented Scheduler for Static and Dynamic Heterogeneity}




\author{Jing Chen}
\affiliation{\institution{Chalmers University of Technology}}
\email{chjing@chalmers.se}

\author{Pirah Noor Soomro}
\affiliation{\institution{Chalmers University of Technology}}
\email{pirah@chalmers.se}

\author{Mustafa Abduljabbar}
\affiliation{\institution{Chalmers University of Technology}}
\email{musabdu@chalmers.se}

\author{Miquel Peric\`as}
\affiliation{\institution{Chalmers University of Technology}}
\email{miquelp@chalmers.se}

\renewcommand{\shortauthors}{J. Chen, P. Noor Soomro, M. Abduljabbar and M. Peric\`as.}

%
\begin{abstract}
With the emergence of heterogeneous hardware paving the way for the post-Moore era, it is of high importance to adapt the runtime scheduling to the platform's heterogeneity.
To enhance adaptive and responsive scheduling, we introduce a 
Performance Trace Table (PTT) into XiTAO, a framework for elastic scheduling of mixed-mode parallelism. The PTT is an extensible and dynamic lightweight manifest of the per-core latency that can be used to guide the scheduling of both critical and non-critical tasks. By understanding the per-task latency, the PTT can infer task performance, intra-application interference as well as inter-application interference. We run random Direct Acyclic Graphs (DAGs) of different workload categories as a benchmark on NVIDIA Jetson TX2 chip, achieving up to 3.25$\times$ speedup over a standard work stealing scheduler. 
To exemplify scheduling adaption to interference, we run DAGs with high parallelism and analyze the scheduler's response to interference from a background process on a Intel Haswell (2650v3) multicore workstation.
We also showcase the XiTAO's scheduling performance by porting the VGG-16 image classification framework based on Convolutional Neural Networks (CNN).
\end{abstract}

%
%


%

%

%
\maketitle

\section{Introduction}

In order to deal with the stringent energy constraints of current IT systems, modern HPC systems are increasingly being architected as heterogeneous platforms. 
Systems may include both static and dynamic sources of heterogeneity. Static heterogeneity sources are those that are fixed at design time, for example, single-ISA cores with different power-efficiency (e.g.~big.LITTLE), or asymmetric ISA systems consisting of processor cores and accelerators (e.g.~CPU/GPU/FPGA platforms). Dynamic sources of heterogeneity are those that arise from runtime reconfiguration. For example, usage of Dynamic Voltage-Frequency Scaling (DVFS)~\cite{dvfs} to tune the performance and efficiency of individual cores or clusters of cores is an example of dynamic heterogeneity. Another example is the usage of cache partitioning to tune cores to the working sets of applications~\cite{iyer-sigmetrics07}. The landscape of heterogeneous configurations creates a challenging scheduling problem. 

To make matters worse, there are many sources of uncontrolled heterogeneity generally called \textit{interference}. Interference refers to performance degradation resulting from shared resources. Interference can occur within a process, e.g.~ resulting from oversubscription of caches and memory bandwidth, or it can occur across processes, e.g.~resulting from time-sharing of a single processing unit in order to run multiple concurrent applications. Obviously, designing scheduling strategies that address all these sources of heterogeneity under a optimization target (performance, energy, etc.) is a complex task. A solution necessarily requires online monitoring that can identify and adapt both to static and dynamic heterogeneity. 

A scheduling solution for multithreaded computations requires an overlying execution model that is flexible and effective for performance portability. Prior research suggests that mixed-mode parallel computations, in which the nodes of a computational task-DAG are themselves parallel computations that can be assigned different amounts of processing resources~\cite{pericas-taco18,chakrabarti-jpdc97,wimmer-spaa11}, are a promising execution model for modern hierarchical and heterogeneous platforms. 
%
An important result from our research~\cite{pericas-taco18} is that greedy scheduling~\cite{blumofe-jacm99} 
is often an undesirable goal as it easily leads to resource over-subscription, particularly with the bandwidth-constrained and cache-constrained nature of modern compute systems\footnote{\url{https://www.karlrupp.net/2013/06/cpu-gpu-and-mic-hardware-characteristics-over-time/}}. This occurs because greedy schedulers will always try to schedule a ready task as long as there are idle processing units, even if the resulting interference results in a global deterioration of performance. 
One way to avoid the problem of resource oversubscription is to aggregate enough resources into a single execution place such that potentially interfering processes are forced to wait. In the context of mixed-mode parallelism this translates into providing the internally parallel tasks with enough resources to avoid interference. 


The XiTAO library\footnote{\url{https://sites.google.com/site/mpericas/xitao}} is an embodiment of this concept. In XiTAO, parallel tasks are scheduled into resource partitions called Elastic Places~\cite{pericas-taco18}. The method has been shown to perform efficiently on homogeneous manycore and NUMA systems. 
To achieve higher power efficiency, however, it is important to make XiTAO aware of heterogeneous platforms. Furthermore, XiTAO traditionally requires the programmer to statically determine the size of the N-core-places. This property limits productivity and performance portability. 

In this paper, we explore and propose schemes to automatically determine resource partitions at runtime. Furthermore, we research how this knowledge can be used to leverage modern single-ISA platforms with both static and dynamic sources of heterogeneity. To this end, we present a scheduler inspired by Criticality-Aware Task Scheduling (CATS)~\cite{chronaki-ics15} and extend it with a performance trace table (PTT) that monitors the system's performance characteristics at runtime. Despite its simplicity, the PTT provides enough information to implement both heterogeneity-aware and interference-free schedules at runtime at minimum cost. Most notably, these features are achieved with no static knowledge about the features of the application and no platform knowledge beyond what can be easily obtained with a tool such as \texttt{hwloc}~\cite{goglin-hal16}.
%

To validate our proposal, we evaluate the scheme using irregular DAGs 
and the VGG-16 neural network on both an Nvidia Jetson TX2 development board featuring an heterogeneous multicore architecture, 
and on an Intel Xeon Haswell platform composed of two NUMA nodes and a total of 20 cores. Our experiments focus on the benefits of criticaltity-aware and interference-free scheduling on both static and dynamic heterogeneity. Sources of heterogeneity include static platform characteristics and dynamic interference episodes, such as co-scheduling of conflicting processes. We conclude by analyzing the scalability of a VGG-16 implementation on the PTT-enabled XiTAO framework.




In summary, this paper contributes:
\begin{itemize}
    \item A heterogeneous scheduler on top of XiTAO and a study of the impact of its main tuning parameters.
    \item A tracing scheme called Performance Trace Table (PTT) that allows to infer the dynamic system heterogeneity.
    \item An in depth evaluation of adaptability of the PTT in the context of interference and architectural heterogeneity.
    \item A showcase implementation of VGG-16 on XiTAO and a scalability analysis.
\end{itemize}
The remainder of this paper is organized as follows. 
Section~\ref{background} describes the background of graph theory which is used to describe computations.  
We present our approach in Section~\ref{approach}. Section~\ref{setup} describes the  experimental setup which is used to evaluate our approach (Section~\ref{evaluation}).
Section~\ref{RelatedWork} describes related work, while Section~\ref{conclusion} concludes the work. 

\section{Task-DAG Scheduling}
\label{background}
\begin{figure}[t]
\centering
\includegraphics[width=0.32\columnwidth]{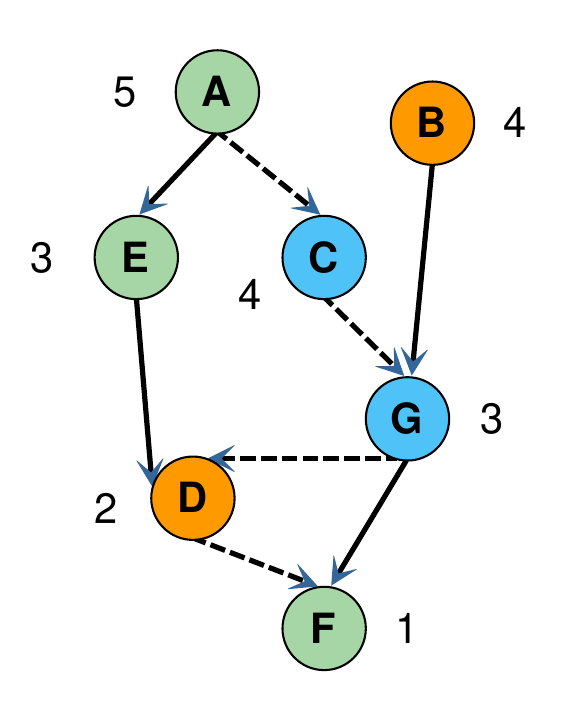}
\caption{An illustration of a DAG with a critical path in dotted lines. Different color represents different kernels. The number of tasks is seven and the critical path has length five. 
}
\label{fig:critical_path}
\end{figure}

This work focuses on computations that can be described as DAGs (Directed Acyclic Graphs). An example of a task-DAG is shown in Figure~\ref{fig:critical_path}.
We define the critical path of a task-DAG as its longest path. 
In Figure~\ref{fig:critical_path}, the critical path is marked with dotted lines whose length is 5. This can be obtained by traversing the nodes $\mathbf{A}$ $\rightarrow$ $\mathbf{C}$ $\rightarrow$ $\mathbf{G}$ $\rightarrow$ $\mathbf{D}$ $\rightarrow$ $\mathbf{F}$. 
Tasks in the critical path are from now on called \textit{critical tasks}, while the others are \textit{non-critical tasks}.
Nodes $\mathbf{B}$ and $\mathbf{E}$ are non-critical tasks in Figure~\ref{fig:critical_path}.
The task-DAG in Figure~\ref{fig:critical_path} also shows the criticality values of each node.
Setting the criticality value in the DAG is achieved by traversing the DAG bottom-up until it reaches the start node(s). Hence, this requires the full DAG to be available before execution. The criticality value is set to be the maximum criticality of children plus 1. Effectively, this results in the first node of the longest path having the highest criticality value. As we can see, task $\mathbf{A}$ has the highest criticality.
This strategy also shows how criticality can be determined at runtime: the task's current criticality value is compared with the parent's criticality value. If the difference is 1, then the task is on the critical path.
Another important concept in this paper is the average DAG parallelism, which we define as $Parallelism = \frac{Number\ of\ total\ tasks}{Number\ of\ critical\ tasks}$. For example, the parallelism in Figure~\ref{fig:critical_path} is 7/5=1.4.
\section{Scheduling for Heterogeneous Systems}
\label{approach}
We begin this section by introducing the main concepts of the XiTAO runtime. We then present the performance trace table (PTT), a data structure that implements an online model of the execution time of each task type. 
Based on the PTT data structure, we then describe our implementation of the performance-based scheduler.


\subsection{The XiTAO library}

XiTAO is a novel runtime for executing mixed-mode computations in which the individual tasks of a task-DAG are themselves parallel computations. These parallel computations are usually data-parallel computations, but any sort of parallel structure is possible. In XiTAO, the individual task is a Task Assembly Object (TAO) and the task-DAG is called the TAO-DAG. For generality, we use the terms task and TAO interchangeably in this paper. 
A TAO type contains a concurrent computation, an internal scheduler, and a \textit{resource width}. 
At runtime, the TAO type is instantiated by providing input arguments to its functionality. 
The resource width denotes how many cores are used to execute a task. The resource width must be a natural divisor of the number of available logical cores in a particular core-cluster, e.g. such as a NUMA domain. The \textit{leader core} is the logical core with the smallest id. Resource partitions assigned to TAOs are composed of consecutive core ids. At execution time, the runtime pins the logical threads to physical cores so that consecutive thread ID's map to cores sharing the same last level cache. 

The XiTAO scheduler implements two queues for each core: a work stealing queue (WSQ) and a FIFO assembly queue (AQ). 
The WSQs store the ready tasks and use random work stealing as a policy for load balancing.
When a ready TAO is fetched from the WSQ and its resource width is determined, pointers to the TAO are then inserted into all AQs representing the resource partition of the TAO. Subsequently, each core asynchronously fetches these pointers and executes the TAO. Note that the resource partitions are irrevocable once assigned. Once the TAO pointers have been inserted into the AQs, the TAO must necessarily execute in the selected set of resources. In other words, all scheduling decisions must happen before the TAO is inserted in the AQs. 
More details about the theory and implementation of XiTAO can be found in~\cite{pericas-taco18}.

\subsection{Performance Trace Table}

\begin{figure}[t]
\centering
\hspace*{-0.5cm}\includegraphics[width=1.1\columnwidth]{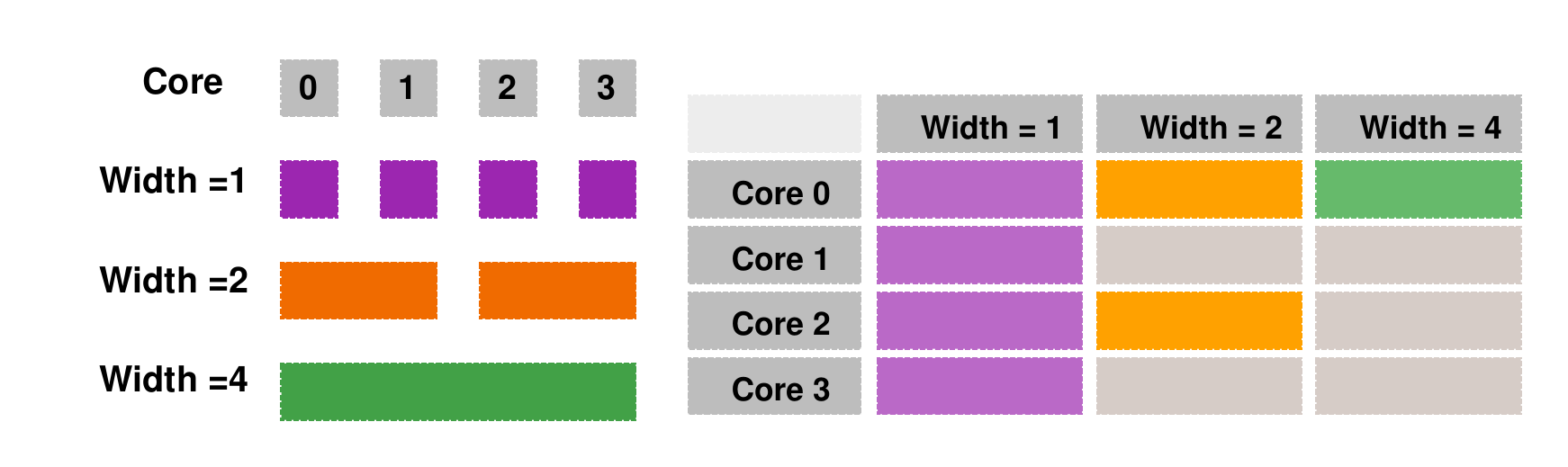}
\caption{Example of a PTT with four cores. The resource width can be 1, 2, or 4.}
\label{fig:time-table}
\end{figure}

Typical scheduling implementations surveyed in Section~\ref{RelatedWork} assume prior knowledge of task loads. However, that is not applicable in our case where the runtime has no prior knowledge of the core type.
In a heterogeneous platform, this work considers the case in which no assumption is made on which type of cores are faster and which kinds of tasks are more suitable for the different core types.
Therefore, to be able to intelligently distribute tasks to the corresponding core type and to dynamically affect the scheduling decisions based on the available resources, we introduce a performance tracer of tasks at runtime and a table (PTT) to model task execution times. 
The table provides an online model of the execution time for each valid combination of \textit{leader core} and \textit{resource width}, \textit{(core id, resource width)}. 
The left part of Figure~\ref{fig:time-table} shows all the scenarios of resource width when the total number of cores seen by the runtime is four.
Therefore, the resource width in this case can be 1, 2 and 4.


The PTT is implemented in the XiTAO runtime.
It is organized as shown in the right of Figure~\ref{fig:time-table}. 
The size of the table is $\mathit{core\_number} \times \mathit{resource\_width\_number}$.
The fields of the table are initialized to 0 that models a zero execution time. This ensures that all configuration pairs will eventually be visited and trained at runtime. 
Due to the decentralized implementation of the scheduler, the table is organized to fit into cache lines where each core only accesses one cache line indexed with core number, hence avoiding false sharing. 
For each entry, the execution time of the pair is temporarily stored.
Then each entry is updated with a weighted time of 1:4, thus, 
the old execution time of the entry occupies 80\% and the new time occupies 20\%.
That is, $\mathit{updated\_value}=\frac{(4 \times \mathit{old\_value}) + \mathit{new\_value}}{5}$.\\
The performance trace table is updated always by the leader core of a task. This simplifies the implementation and reduces cache migrations. 
This also means that every core can have a model value of the task with $\mathit{width}=1$ but only every fourth core will have a model value of the task with $width=4$.
As shown in Figure~\ref{fig:time-table}, in the case of width=1 (purple field), each core is the leader for its own partition.
For the case width=2 (orange), the leading core 0 (2) handles the resource partition containing cores 0 (2) and 1 (3).
By restricting the leader core to update the PTT, a potential skew of the model may result, since the leader may not provide the most accurate record of the execution time of that TAO, i.e. it could have had the least or the most amount of work. This is because these workers enter and exit the execution of the TAO asynchronously~\cite{pericas-taco18}. 
However, the weighted average that is used to updated PTT values ensures that the impact of imbalance will be limited.
Although averaging results in an additional read of the table, the table size is small and it is very important to be resilient to divergent measurements as this table is the key point of scheduling decisions. While the impact of this feature requires a detailed study, our experiments so far do not provide evidence that the potential imbalance is problematic.

This implementation of tracing execution history requires as little information as the number of cores and their distribution into core-clusters with shared caches. The cores simply update the corresponding index, independent of its resource type, and thus create a model of performance.
In the other words, no matter what core-types a platform has, be them big or little, the performance of these cores will be reflected by PTT values.
This is beneficial not only for portability and potentially functional-heterogeneity, but also for temporally added heterogeneity such as DVFS caused by heat variations, or interference caused by other tasks and/or uncontrollable system activities such as background processes or interrupts. 

\begin{figure}[t]
\centering
\hspace*{-0.7cm}\includegraphics[width=1\columnwidth]{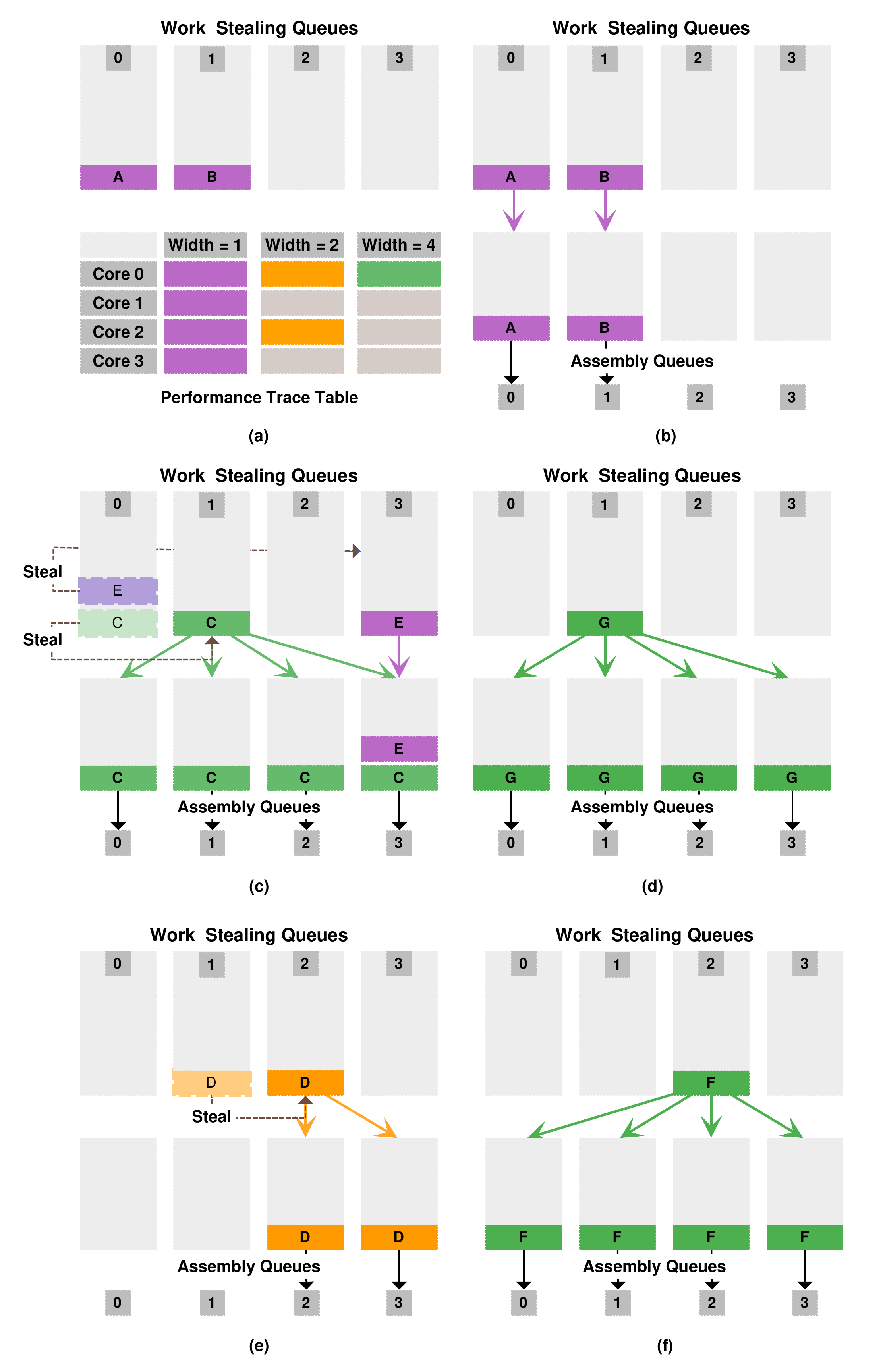}
\caption{The scheme of performance-based scheduler. It includes seven tasks, distributing into different work stealing queues with the same resource width 1.}
\label{fig:perf_sched}
\end{figure}


\subsection{Performance-based Scheduler}

Based on the implementation of the performance trace table, we develop a heterogeneity-unaware scheduler with the goal of optimizing performance. To achieve this, we follow the basic strategy of CATS~\cite{chronaki-ics15} and extend it with our heterogeneity-unaware methodology. This scheduler is named performance-based scheduler. 
The main feature of the performance-based scheduler is the ability to find the optimal cores and resource width using the optimal values depicted by globally searching the PTT.

The operation of the performance-based scheduler follows four steps.
Figure~\ref{fig:perf_sched} is an example of the scheduler implementation based on the task-DAG shown in Figure~\ref{fig:critical_path}.
Firstly, we fetch the ready tasks from our application task-DAG. 
These tasks are then inserted into a particular work stealing queue according to the default policy or to programmer annotations. 
When the task reaches the head of the WSQ, it is permitted to be executed locally or randomly stolen. The task's priority is checked to detect whether the task belongs to the critical path, and thereafter, decide the execution policy.
Note that the criticality of all the tasks of a task-DAG decides the priority.

If a task is critical, e.g. task $\mathbf{A, C, G, D}$ or $\mathbf{F}$, we globally search the performance trace table to find the optimal pair of core and resource width for such task. Global search means that all the entries of the PTT of the particular TAO type are checked to find the value that globally minimizes $\mathit{exec\_time} \times \mathit{resource\_width}$. The goal of this operation is to find the pair of core and resource width that globally minimizes the system's occupation of resources, understood as the product of resources and execution time. Alternative optimization strategies are also possible. For example, a system trying to minimize the energy consumption would instead find the best pair that minimizes energy per task. 

The overhead of the global search operation is low since the number of entries in the PTT is only $2 \times N - 1$ for each NUMA node consisting of $N$ cores.
In this way, we guarantee that all the critical tasks are executed on faster cores, obtaining better performance and minimizing interference.
For non-critical tasks, however, only the appropriate resource width for the corresponding core is determined from the PTT when the TAO is scheduled for execution. The goal of this policy is to reduce interference across non-critical tasks. 
In summary, critical tasks search the PTT globally to improve performance and reduce interference, while non-critical task just search the current core's entries in the PTT with the goal of avoiding interference. 

Initial tasks have no parents and therefore it is not possible to determine their criticality. In the current implementation they are treated as non-critical tasks: the PTT is not globally searched, but we still try to find a good resource width.
For example, we can see that from Figure~\ref{fig:critical_path} that task $\mathbf{A}$ and $\mathbf{B}$ are initial tasks running on threads 0 and 1. They are scheduled to cores 0 and 1 with resource width 1, respectively, according to the best PTT resource width. 
At this point the two tasks are inserted into the corresponding assembly queues, as shown in Figure~\ref{fig:perf_sched} (b).

After a task finishes, the \textit{commit-and-wake-up} stage will check if any dependent tasks are critical. 
If the child task's criticality is less than the parent's by 1, then the task is determined to be critical.
For example, in Figure~\ref{fig:critical_path}, task $\mathbf{C}$ is the child task with criticality value 4 of task $\mathbf{A}$ with criticality value 5.
To this end, after completing the execution of task $\mathbf{A}$, core 0 could wake up another task which is dependent on task $\mathbf{A}$.
In Figure~\ref{fig:perf_sched} (c), task $\mathbf{C}$ and task $\mathbf{E}$, which are the children of task $\mathbf{A}$, are woke up sequentially.

However, if there are no ready tasks in the work stealing queue, the core will try to steal a task from other cores' work stealing queues. 
Since the work stealing queues of core 1, 2 and 3 are empty after completing task $\mathbf{B}$, they can steal tasks from other work stealing queues.
For instance, core 1 and 3 steal task $\mathbf{C}$ and $\mathbf{E}$ from the work stealing queues of core 0 respectively.
As a critical task, task $\mathbf{C}$ globally searches the performance trace table and finds the optimal pair (0,4), i.e.~the leader core is 0 and the resource width is 4, and then this task is distributed into the assembly queues of cores from 0 to 3.  
Task $\mathbf{E}$ is a non-critical task, it searches only the entry for core 3 in the performance trace table for a optimal width.
As resource width=1 is the best choice for core 3, the task $\mathbf{E}$ is distributed into the assembly queue of core 3, as Figure~\ref{fig:perf_sched} (c) shows.
Since its parent tasks $\mathbf{B}$ and $\mathbf{C}$ have completed, task $\mathbf{G}$ is woken up by core 1.
By searching the performance trace table, it finds out that the pair (0,4) is the best one, then core 1 distributes the task into assembly queues from 0 to 3 (see Figure~\ref{fig:perf_sched} (d)).
Then, as shown in Figure~\ref{fig:perf_sched} (e), the critical task $\mathbf{D}$ is woken up by core 1 but stolen by core 2 and the optimal pair for it is determined to be (2,2).
Core 2 complete the task $\mathbf{D}$ and then wake up its final children and the critical task $\mathbf{F}$ with the best configuration (0,4) in performance trace table, as shown in Figure~\ref{fig:perf_sched} (f).
\section{Experimental Setup} \label{setup}

\subsection{Evaluation Platforms}
The benchmarks herein are evaluated on two platforms. From the static heterogeneous family, we use a NVIDIA Jetson TX2 development board, featuring a dual-core NVIDIA Denver 2 64-bit CPU, a quad-core ARM A57 Complex (each with 2 MB L2 cache) and an NVIDIA Pascal Architecture GPU with 256 CUDA cores. Both the Denver 2 and the A57 cores implement the ARMv8 64-bit instruction set and are cache coherent. For the purpose of this work, we consider only the two ARMv8 cores, and leave GPU scheduling as future work. 
On the homogeneous side, an Intel 2650v3 (code-named "Haswell") based platform is used to evaluate the effect of interference while scheduling Random DAGs, and to evaluate the behavior of XiTAO when executing the Image Classification network (VGG-16)~\cite{simonyan-arxiv14}.

\subsection{Random Directed Acyclic Graph}
\subsubsection{Kernels}

We generate random DAGs to evaluate the properties of the PTT-enhanced scheduler. The random DAGs are based on a mix of different kernel types.  
When selecting the kernels, the priority is to achieve different characteristics in terms of memory-intensiveness (streaming), cache-intensiveness (i.e. data reuse) and compute-intensiveness. The following three kernels are selected for this purpose.

\textbf{Matrix Multiplication.} A \textit{matrix multiplication} kernel is created for the \textit{compute-intensive} property. We implement a matrix multiplication that achieves parallelism by ensuring that the writing of output data is done to separate cache lines for each thread while still sharing the input data. 

\textbf{Sort.} 
For the \textit{data reuse} property, a \textit{quick sort} and \textit{merge sort} kernel combination is selected. 
This kernel first splits the input array into chunks and performs in-place sorting with quick sort before carrying out two levels of merge sort, effectively reusing the data within the kernel. This kernel has a maximum parallelism of four. 

\textbf{Copy.} 
Finally, a \textit{copy} kernel handling large inputs is implemented for the \textit{streaming} property. 
This kernel reads and writes large portions of data to memory, effectively creating a streaming behavior where the kernel has to access the main memory continuously. Each core copies a subset of the data.

For each kernel, we select the appropriate working set size corresponding to the desired behavior. 
For the matrix multiplication kernel, we choose a 64 $\times$ 64 matrix.
For the sort kernel, we choose a 262KB input array, taking up a total space of 524KB due to double buffering, effectively fitting iton the L2 caches. 
Finally, the copy kernel uses a 16.8MB array, taking up a total space of 33.6MB, which is much larger than the space of the L2 caches. 

\subsubsection{DAG construction}
To properly evaluate the performance of our scheduler, randomized DAGs composed of random selections of these three kernels are implemented. 
By tuning the parameters, it is possible to achieve different degrees of average parallelism and thus generate different scheduling scenarios.

To generate a suitable randomized DAG, a set of configuration parameters are used, similar to the generation of DAGs by Topcuoglu et al.~\cite{topcuoglu-tpds02}. 
The first parameter is the number of tasks of each kernel. This is useful to choose which kernel should be most prominent in the DAG. 
The second parameter is the average width of the DAG. This is used to obtain the desired level of parallelism.
The last configuration parameter, the edge rate parameter, determines the average amount of connected edges a task has, which also affects the parallelism of the DAG. 
A seed value is used to manipulate the randomization to recreate a different DAG several times for comparison.

The DAG generation algorithm produces a DAG in three steps. 
The first step generates the shape (nodes and edges) of the DAG. 
This step is separated from the TAO creation step in order to get proper memory utilization and data reuse (see below).
The second step consists of allocating memory and deciding which tasks are reusing data. 
To achieve data reuse between nodes, we maintain a vector for every kernel where each index in the vector represents a memory location. 
Initially, the size of the vector is zero. 
For every node, we search its predecessors for a node number matching any of the numbers in the vector. 
If a matching number is found, it is replaced by our current node number and the index to the location is saved in the node. 
If we cannot find a match, a new entry is created in the vector with the unique node number and that index is saved in the node instead. 
The size of the vectors is then used for allocation of memory and each node will have a designated data location. 
The memory is allocated this way to maximize data reuse between tasks of the same kernel while guaranteeing isolated data execution when tasks are run in parallel.
The final step is to traverse the nodes and spawn the corresponding tasks and edges between them, thus effectively creating the DAG in the XiTAO format. 

\subsection{Image Classification}

\begin{figure}[t]
\centering
\includegraphics[width=0.9\columnwidth]{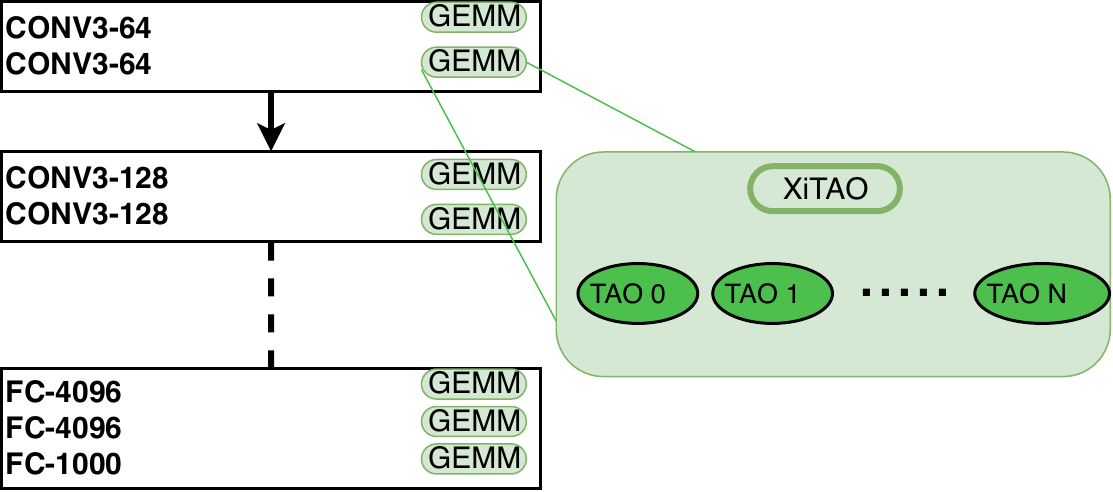}
\caption{Architecture of VGG-16. here CONVX-Y represents X-D filter and Y Channels of convolutional layer respectively}
\label{fig:vgg}
\end{figure}

To demonstrate the behavior of the performance-based scheduler in XiTAO, we port the VGG-16~\cite{simonyan-arxiv14} image classification model from the Darknet framework~\cite{redmon-cvpr16}. The application uses a 16 layered deep convolutional neural networks (CNNs) to classify an image using a pre-trained model. Each convolutional (CONV) and fully-connected (FC) layer implements GEneral Matrix Multiply (GEMM)
that takes most of the computation time. 
Figure \ref{fig:vgg} shows the XiTAO implementation of VGG-16. In VGG-16, input size varies as the network progresses. For example, the convolutional layer iterates over a minimum 64 channels to a maximum of 512 channels. Therefore, in the XiTAO implementation we partition the work among TAOs. The number of TAOs in each layer depends on the number of \textit{channels} and \textit{block\_length}. The parameter \textit{block\_length} refers to the number of channels assigned to each TAO, which is tuned at runtime. Each TAO performs parallel GEMM with the number of threads equal to the \textit{width} of TAO. Note that the \textit{width} is dynamically determined by the XiTAO scheduler. Since there are no loop carried dependencies inside the layer we benefit from two levels of parallelism in the XiTAO implementation. However, each layer is dependent on the previous layer, we therefore synchronize all TAOs at the end of each layer.

\section{Performance Evaluation}
\label{evaluation}
The outline of the evaluation is as follows: we first analyze the impact of using the performance-based scheduler versus the homogeneous scheduler (i.e.~the base random work stealing algorithm as implemented in XiTAO~\cite{blumofe-jacm99}) that is both unaware of the hardware and of the ongoing performance state modeled by the PTT. We study Random DAGs consisting of a mixture of kernels (MatMul, Sort, Copy) to cover the spectrum of real DAGs as much as possible. We also port the Darknet VGG-16 code to XiTAO and compare it to the base CPU implementation to assess the enhancements on Convolutional Neural Networks that are of high relevance to contemporary applications.  


\subsection{Comparison with Homogeneous Scheduler}
\begin{figure}
\centering
\subfigure[Performance-based Scheduler]{\label{fig_hetero} \includegraphics[width=0.45\textwidth]{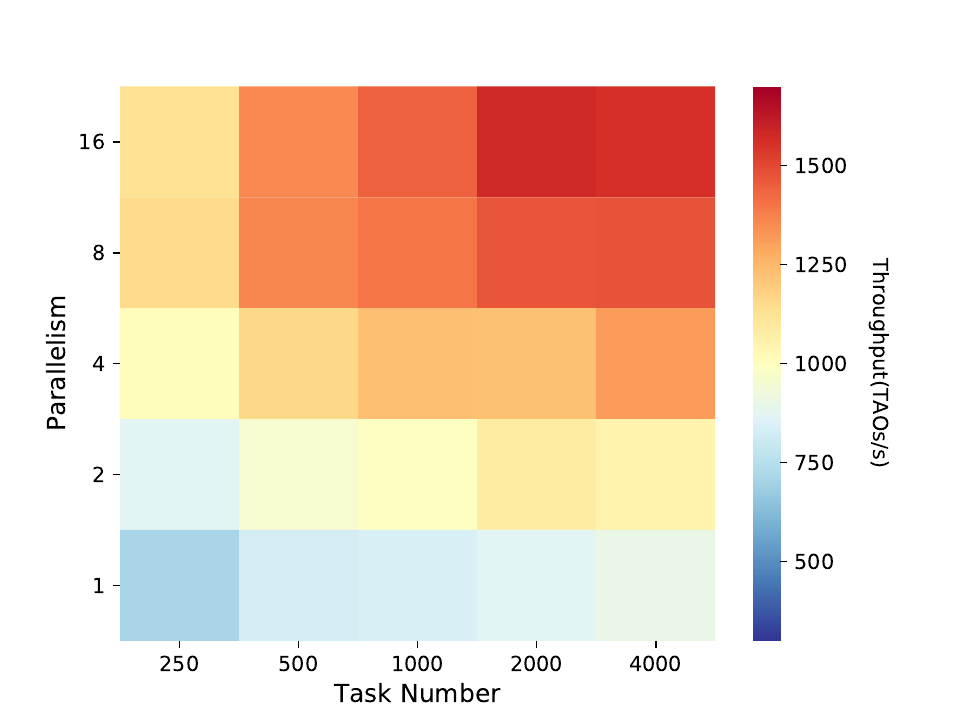}}
\subfigure[Homogeneous Scheduler]{\label{fig_homo}
\includegraphics[width=0.45\textwidth]{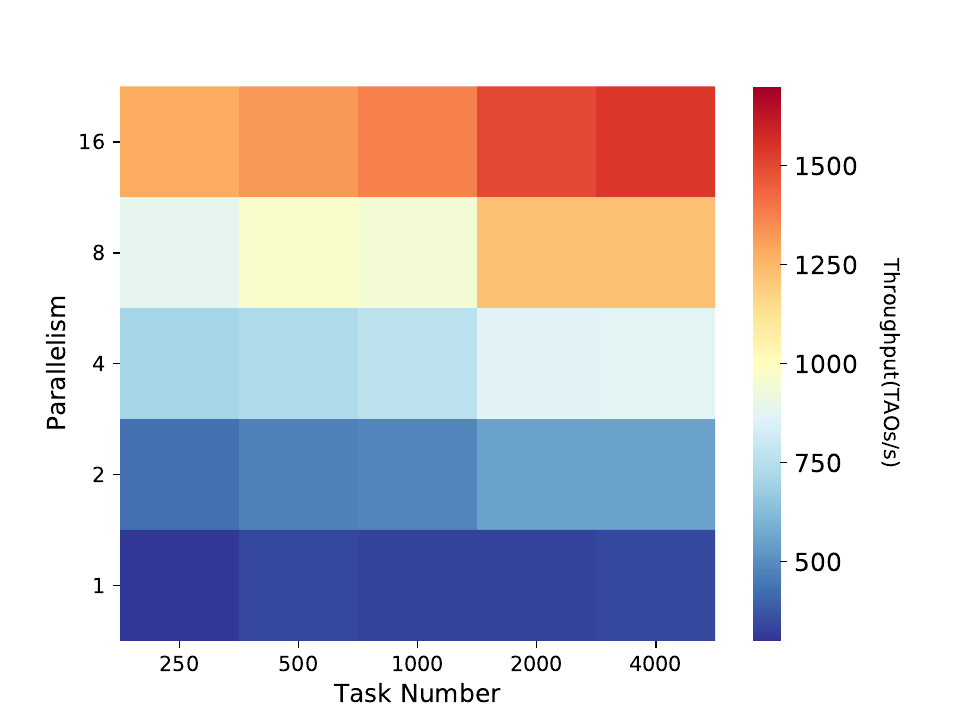}}
\caption{The performance impact over parallelism and number of TAOs and the performance comparison between performance-based scheduler and homogeneous scheduler.} 
\label{fig_hetero_homo}
\end{figure}

Figure~\ref{fig_hetero_homo} denotes a heatmap per each described scheduler executing between 250-4000 tasks (X-Axis) on random DAGs with a parallelism between 1-16 (Y-Axis). The underlying random DAG is a combination of the aforementioned kernels with equal proportions. In the most challenging case with low task count and parallelism (tasks=250, par=1), we observe that the temperature of the performance-based scheduler (depicted by Figure~\ref{fig_hetero}) is at least twice higher. The additional ingredient of scheduling critical tasks on the high performing cores (mainly \textit{Denver} cores in this case) and the ability to dynamically tune the resource width renders this scheduler superior even with no external task-DAG parallelism.
The throughput is higher across the table except for a few cases of very high parallelism that pose almost no challenge on scheduling decisions. Another interesting yet expected observation is that the number of tasks plays a negligible role on the performance of the homogeneous scheduler, whereas the throughput of the counterpart is a factor of both axes. A twofold increase in the number of tasks provides twice the amount of PTT training data. This directly reflects on the performance by improving the quality of the dynamic, PTT-based choices. In addition, a higher degree of parallelism (on the Y-Axis) permits a better utilization of the resources.   

\begin{figure*}[!t]
\centering
\subfigure[Performance-based Scheduler]{\label{fig_line_hetero} \includegraphics[width=0.40\textwidth]{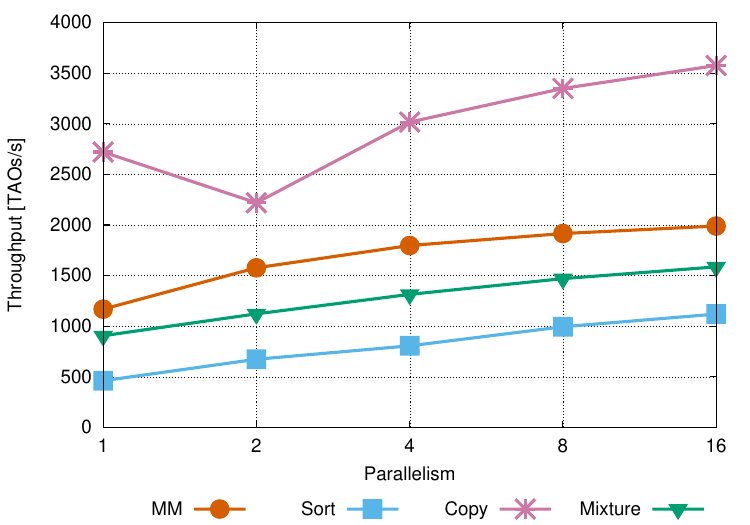}}
\subfigure[Homogeneous Scheduler]{\label{fig_line_homo}
\includegraphics[width=0.40\textwidth]{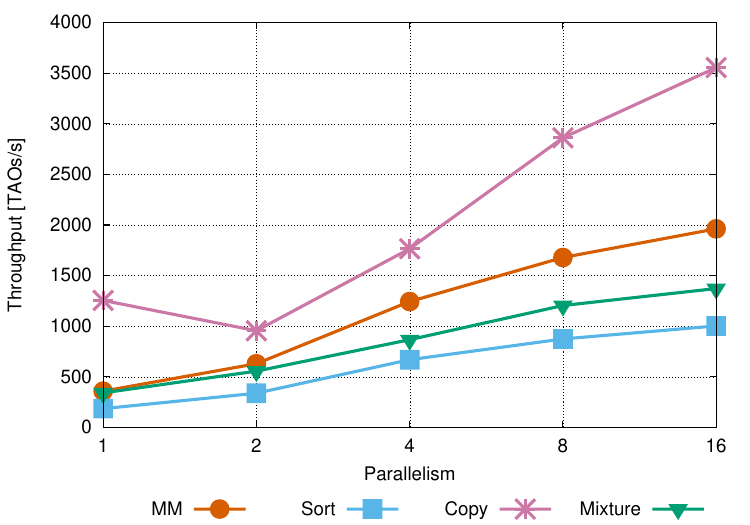}}
\caption{The performance impact over parallelism and kernels and the performance comparison between performance-based scheduler and homogeneous scheduler.} 
\label{fig_line_hetero_homo}
\end{figure*}

\subsection{Performance Impact of Kernels}
\begin{figure}[h]
\centering
\includegraphics[width=0.41\textwidth]{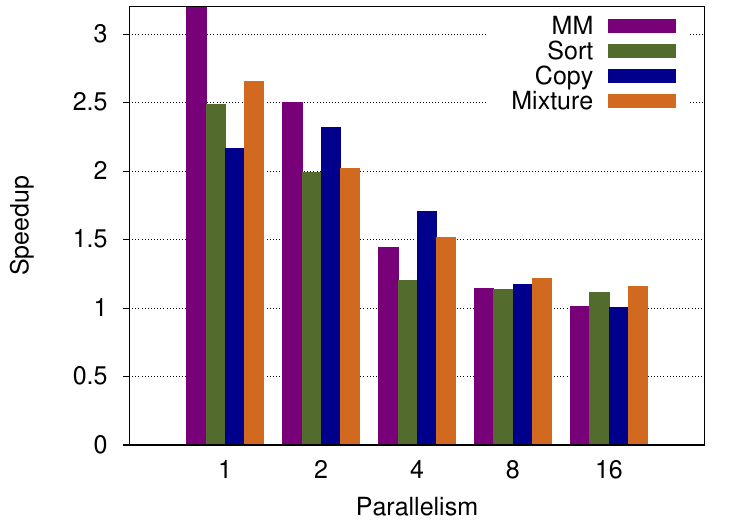}
\caption{The performance speedup when comparing performance-based scheduler with homogeneous scheduler with different parallelism. In this figure, the number of tasks we use is same as Figure~\ref{fig_line_hetero_homo}.}
\label{fig_speedup_linechart}
\end{figure}

As highlighted before, we select three different kernels \textit{matrix multiplication, sort, copy} with different characteristics. 
It is therefore important to evaluate the performance impact of such kernels and their mixture while varying parallelism. Figure~\ref{fig_line_hetero_homo} compares the throughputs of the performance-based scheduler and homogeneous scheduler on the Jetson TX2 platform for various degrees of parallelism.
Besides the higher throughput achieved by the performance-based scheduler especially for lower parallelism, it exhibits a greater stability across the X-axis, which is an essential attribute that suggests that the hardware is being efficiently utilized and that the scheduler is less sensitive to parallelism constraints. For a concrete quantification of performance gains, Figure~\ref{fig_speedup_linechart} shows the speedup achieved using the performance-based scheduler over the homogeneous scheduler. 
In this case, we use 4000 tasks for each kernel (i.e., \textit{matrix multiplication}, \textit{sort} and \textit{copy}), and a Random DAG that contains a mixture of even number of tasks/kernel that sum up to 4000.  
We can see that our performance-based scheduler generally runs faster than homogeneous scheduler.
Specifically, it has significant speedup when the parallelism is low.
For a parallelism of 1, it achieves 3.3$\times$ of the throughput when compared with the homogeneous scheduler in \textit{matrix multiplication}.
The speedup for \textit{sort}, \textit{copy} and the mixture of three kernels are 2.5$\times$, 2.2$\times$ and 2.7$\times$ for the same parallelism, respectively.
When the parallelism increases, the speedup of performance-base scheduler compared with homogeneous scheduler decreases, but we still have better performance than homogeneous scheduler. 
For scenarios of high parallelism, criticality-aware scheduling has little impact on performance. Instead, in such scenarios it is important to try to avoid interference due to resource oversubscription. These results highlight how the PTT can be used to select appropriate resource width to avoid oversubscription. This is particularly visible for the case of the \textit{sort} kernel, which relies on good usage of cache capacity. Scheduling too many \textit{sort} kernels in parallel leads to oversubscription and performance degradation. Classical heterogeneity aware schedulers such as HEFT~\cite{topcuoglu-tpds02} or CATS~\cite{chronaki-ics15} are not able to address such scenarios.

\subsection{Process Interference}
\begin{figure*}[t!]
\centering
\subfigure[Dynamic migration of processes in response to PTT spikes during interference.]{\label{fig:interference} \includegraphics[width=0.40\textwidth]{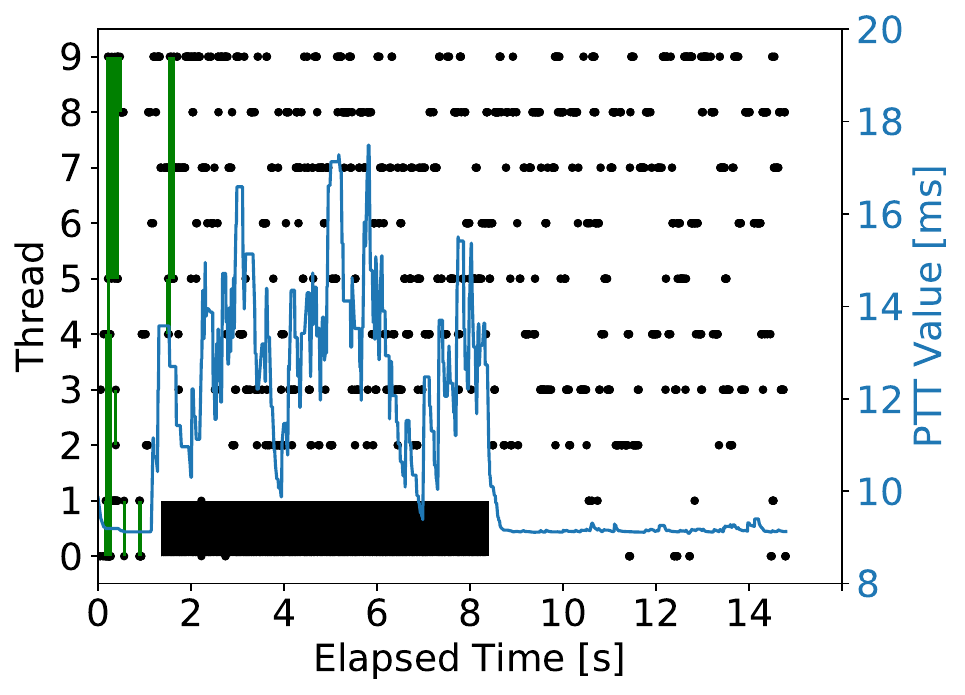}}\hspace{1cm}
\subfigure[The scheduler's behavior when there is no interference]{\label{fig:no_interference}
\includegraphics[width=6.67cm, height=5.1cm]{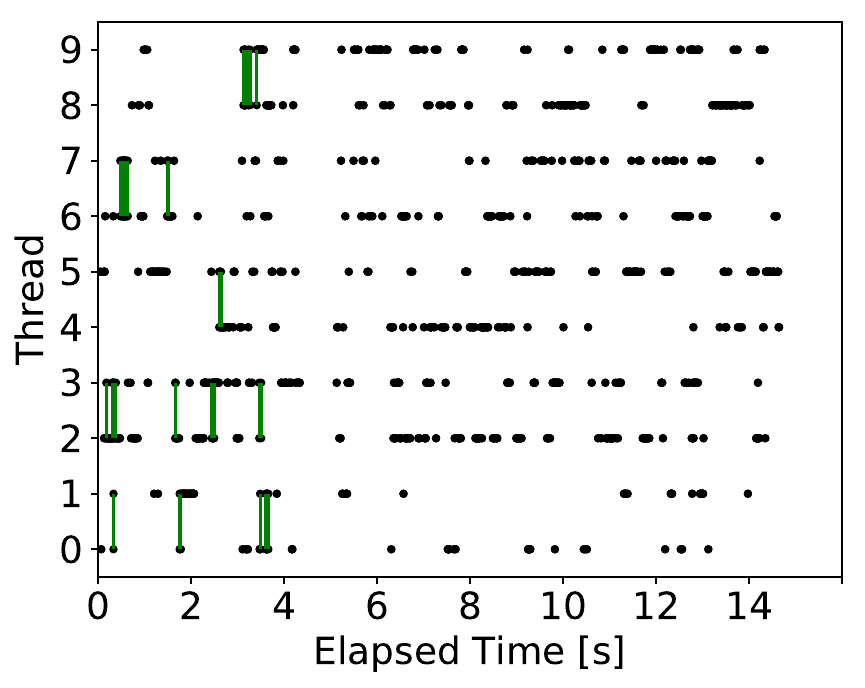}}
\caption{The effect of interference on PTT scheduling of critical tasks.} 
\label{fig:interference_analysis}
\end{figure*}
One of the remarkable advantages of using PTT is maximizing performance via minimizing the side effects of interference. This feature is especially important since it is highly anticipated that user or kernel level resources are shared. Figure~\ref{fig:interference} depicts the response of the XiTAO performance-based scheduler to running a background parallel process, in this case a chain of MatMul DAGs, alongside a highly parallel random DAG. The black dots represent the time-stamp at which the threads start executing TAOs. A vertical green line shows the resource partition used to execute the TAO. While bootstrapping the PTT, a few width choices are attempted. At the point of interference (i.e. in Figure~\ref{fig:interference}), we show the PTT value at (width=1,core=1). Other relevant values are dropped for brevity. Due to the jitters in PTT values, the scheduler automatically selects cores from (2-9) for executing the critical tasks. Cores (0-1) are still selected under typical circumstances according to Figure~\ref{fig:no_interference}. Shortly after the interference event, the scheduler recovers to normal operation yielding a marginal wall time difference across the two experiments. Note that non-critical task continue to be executed on cores with interference, as long as these cores succeed in stealing tasks. This is important so that the PTT is continuously updated to reflect the status of the system. 


\subsection{ImageNET Classification}
\begin{figure}[t]
\centering
\includegraphics[width=0.37\textwidth]{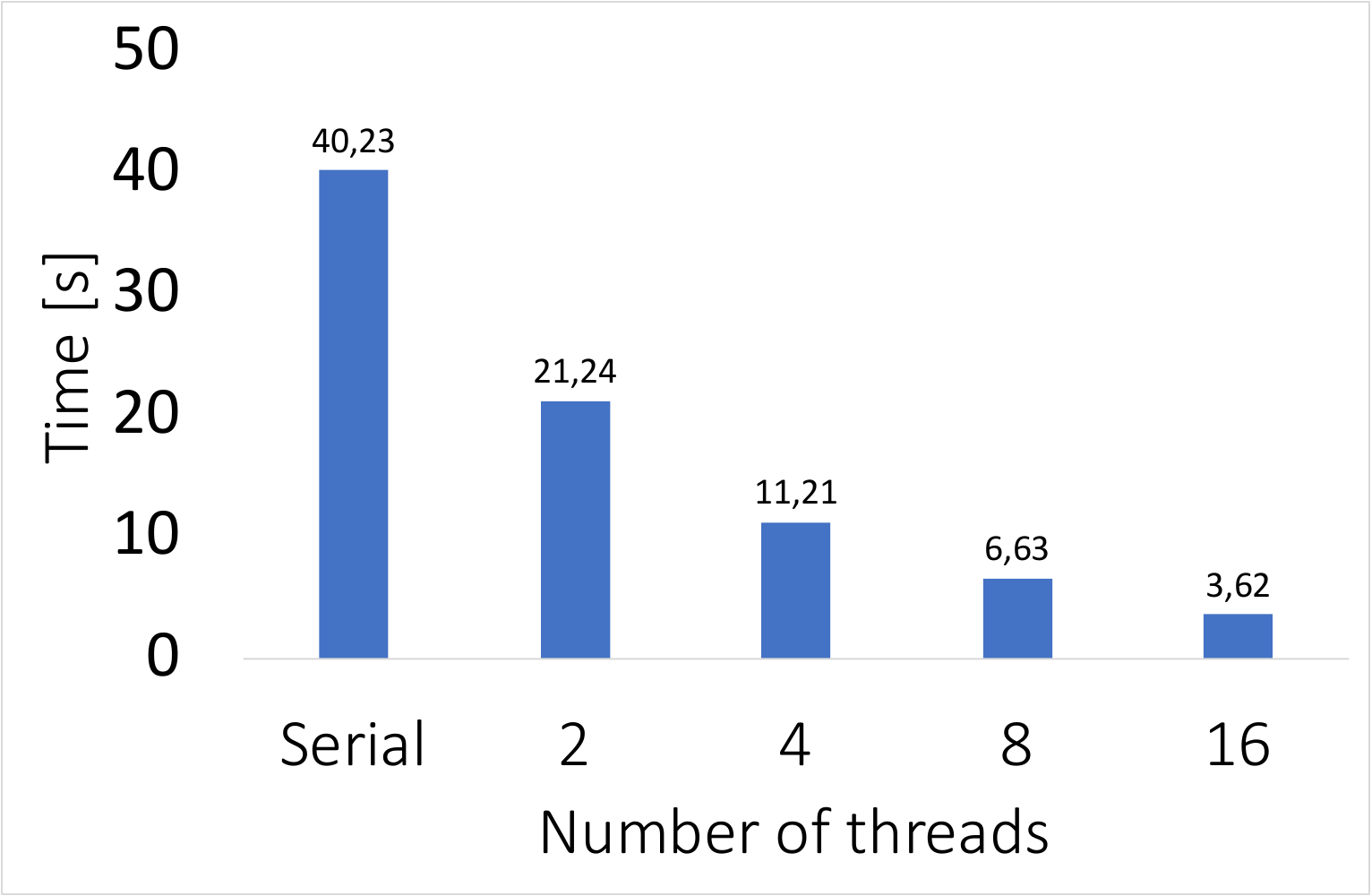}
\caption{Performance of CPU \textit{GEMM} on XiTAO VGG-16 with variable number of threads}
\label{fig:ptt_vgg}
\end{figure}

\begin{figure}[t]
\centering
\includegraphics[width=0.4\textwidth]{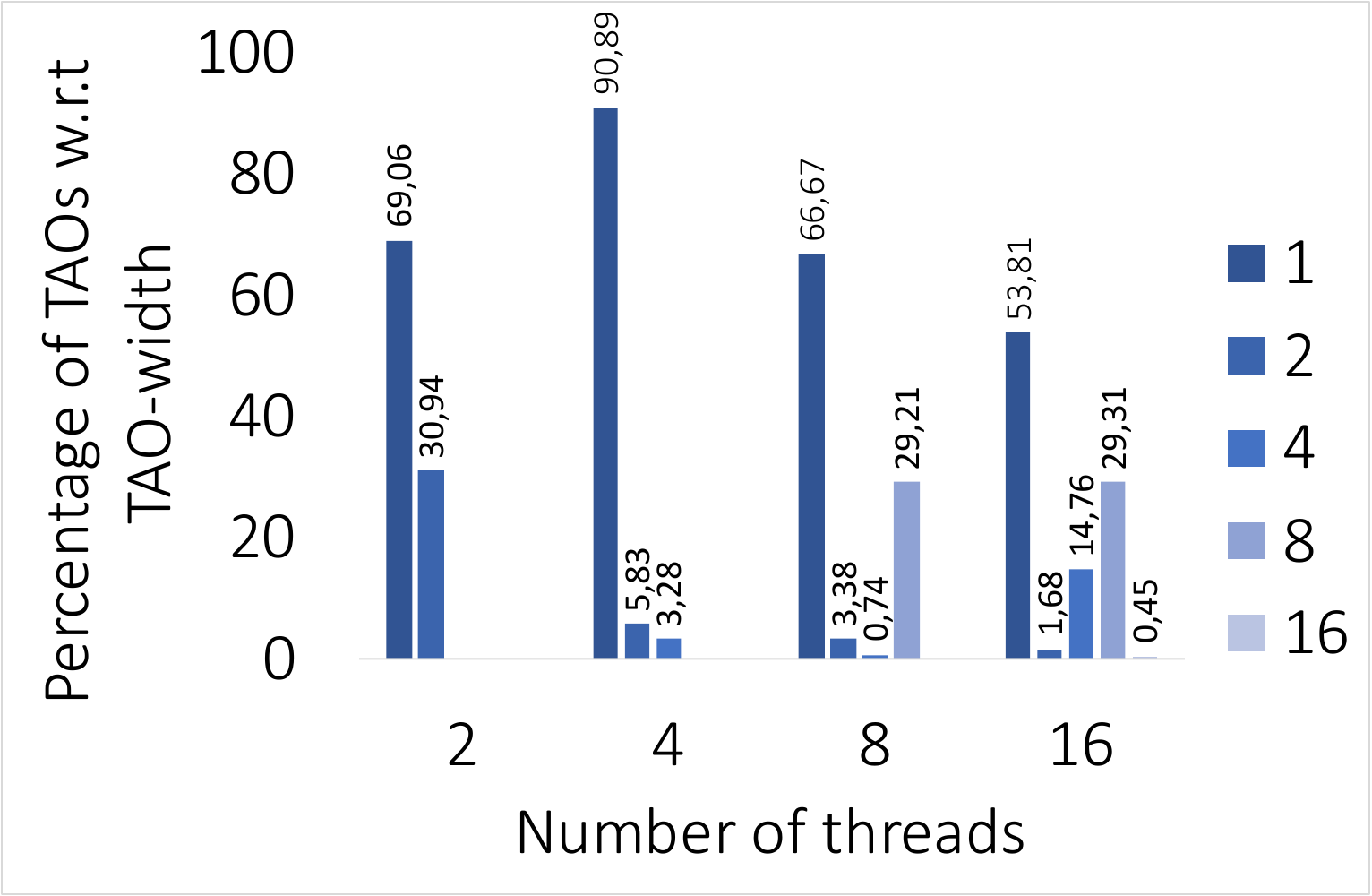}
\caption{Percentage of TAOs scheduled with corresponding TAO width by PTT}
\label{fig:vgg_hist}
\vspace*{-0.2cm}
\end{figure}

Figure~\ref{fig:ptt_vgg} depicts a strong scalability study of the performance of the XiTAO version of the VGG-16 code for predicting a predefined image class by multiple convolutions of a crop layer (1024 x 1024) converted to matrix (512 x 512 x 3). The study is to assess the scheduling performance of the conventional fork-join application class with minimal effort. It is carried out on a dual-socket Intel Haswell platform. 
It is worthwhile noting that XiTAO reorders threads to ensure data locality, since the core ids are not always laid out continuously, as is the case in both this platform and the Jetson TX2 platform. The scheduler still exhibits 0.69 parallel efficiency compared to the serial performance, even though there is no criticality notion to this experiment, i.e., all tasks are marked non-critical.
Figure~\ref{fig:vgg_hist} shows the number of TAOs scheduled with corresponding widths. During execution, PTT chooses the best width to schedule a TAO. For example, in the case of running VGG-16 with 8 threads, \textit{67\%} of TAOs are scheduled with \textit{width = 1} and \textit{30\%} TAOs are scheduled with \textit{width = 8}, indicating that these widths lead to the best speed-up. The approximately linear speedup shown in Figure~\ref{fig:ptt_vgg} combined with the PTT-assisted choices of widths as in Figure~\ref{fig:vgg_hist} demonstrate how negligible the resource tuning overhead is especially if compared to the potential performance gains prescribed by this paper. 




\section{Related Work}
\label{RelatedWork}
\subsection{Scheduling in Heterogeneous Environments}\label{sec:hertrosched}
Task scheduling on a heterogeneous platform, contrary to a homogeneous platform, includes the problem of assigning the appropriate tasks to the most suitable cores. Most multicore scheduling approaches today assume equal performance. For example, dynamic scheduling techniques such as work-stealing or work-sharing do not consider the individual performance of cores.

Scheduling DAGs on heterogeneous multicores is a well studied problem in the context of single-threaded task-DAGs~\cite{topcuoglu-tpds02,cheng2010,chronaki-ics15,chronaki-tpds17,Koufaty,VanCraeynest2012}. These schemes either assign a ranking to each tasks based on the critical path and then assign more critical tasks to faster cores~\cite{topcuoglu-tpds02,cheng2010,chronaki-ics15,chronaki-tpds17}, or they compute a best fit between tasks and cores and then schedule appropriately~\cite{Koufaty,VanCraeynest2012}. In this study we extend upon ideas introduced in CATS~\cite{chronaki-ics15}.
In the following paragraph we describe CATS along with HEFT~\cite{topcuoglu-tpds02}, a classical heterogeneous scheduler, and Bias Scheduling~\cite{Koufaty}.
HEFT is a static scheduling method for heterogeneous task scheduling proposed by Topcuoglu et al.~\cite{topcuoglu-tpds02}. The HEFT algorithm consists of ranking the tasks of a DAG in order of longest path to finish and then assigning the highest-ranking tasks to the core that will minimize the overall finish time. An analysis of the DAG is done to calculate the execution time and communication cost of each node and edge before the tasks can be ranked. The tasks are then placed in a queue where the scheduler picks the top task and calculates which core will be able to finish this task earliest using insertion-based scheduling. 
%
%

 CATS is a dynamic scheduling approach where no prior knowledge about the execution time of the tasks is assumed~\cite{chronaki-ics15,chronaki-tpds17}. 
 Instead, CATS solely uses the number of successors to find the critical path. The critical path is then put in a critical queue. Tasks from the critical queue are scheduled on high-performance cores and tasks from the non-critical queue are scheduled on lower-performance cores. 
 In~\cite{chronaki-ics15}, Chronaki et al.~introduce the dynamic Heterogeneous Earliest Finish Time (dHEFT) algorithm as a reference to evaluate CATS. dHEFT uses the same principles as HEFT but instead of knowing the load of tasks prior to scheduling, discovers them at runtime. 

Finally, Bias Scheduling~\cite{Koufaty} is a proposed method for single-ISA heterogeneous multicore processors that tracks how different kinds of tasks perform on each core. The main idea is to categorize tasks into two groups: Tasks gaining large speedup by running on a big core compared to a LITTLE core and tasks gaining modest speedup by running on a big core. The speedup is approximated by accessing hardware counters for stall cycles. Tasks are then scheduled on big cores if they provide large speedup and on LITTLE cores if the speedup would be modest. 

While all these schedulers can improve the execution time of task-DAGs in which tasks have diverse behaviors, they have a few limitations. First, none of them is able to avoid resource over-subscription and provide interference-free execution. And second, all of them are based on the notion of only two static performance classes, i.e.~big and LITTLE. In practice, sources of heterogeneity are diverse, hence performance needs to be tracked and modeled on a per-core basis. Our proposal can model the performance of all cores and, furthermore, thanks to its reliance on XiTAO it can exploit this information to provide interference-free scheduling.


\subsection{Management of shared resources}

One of the main goals of XiTAO is to provide a good solution for shared resource contention. 
The focus of XiTAO is on multithreaded computations. Prior work on resource contention has mostly 
focused on multiprogrammed workloads, i.e.~multiple single-threaded workloads running in parallel. 
Both software-based scheduling~\cite{zhuravlev-asplos10} and hardware-based partitioning~\cite{qureshi-micro06,iyer-sigmetrics07} 
approaches have been proposed to address issues related to cache and memory sharing. Multiprogrammed
workloads are less challenging in the sense that the different tasks do not have any dependencies. 
Hence, XiTAO addresses a more problematic case in which scheduling and resource partitioning decisions 
can negatively impact the applications execution time, particularly if bad decisions are taken concerning the
critical path of the application. In the context of structured parallelism, such as divide-and-conquer, 
one scheduler that targets constructive sharing is the parallel depth first (PDF) scheduler~\cite{blelloch-jacm99,chen-spaa07}. 
%
Other runtime systems that support mixed-mode parallelism have been proposed by Wimmer et
al.~\cite{wimmer-spaa11} and by Sbirlea et al.~\cite{sbirlea-europar15}. However, they focus only on 
restricted forms of parallelism, such as divide-and-conquer~\cite{wimmer-spaa11} or worksharing constructs~\cite{sbirlea-europar15}. 

\section{Conclusion} \label{conclusion}
We have introduced a performance-based scheduler for heterogeneous architecture that leverages online monitoring on top of the XiTAO runtime. The presented scheduler improves task execution throughput and latency, and it provides interference-free execution and task migration in the event of process interference. All these features allow our proposed scheduler to adapt to next generation heterogeneous systems with shared resources. We evaluated the performance-bsaed scheduler on random DAGs and compared it to the homogeneous counterpart. 

A future direction of this work is to include GPU kernels as part of the scheduling strategy, fully utilizing the Jetson TX2 chip by offloading GEMM kernels of the VGG-16/Random DAGs benchmark to the underlying Pascal GPU. 
We also intend to study the interaction between non-critical tasks and the PTT in terms of updating resources and locality. 

\section{Acknowledgment} \label{ack}
The computations/data handling were enabled by resources provided by the Swedish National Infrastructure for Computing (SNIC) at C3SE partially funded by the Swedish Research Council through grant agreement no. 2016-07213. The research leading to these results has received funding from
the European Union’s Horizon 2020 Programme under the LEGaTO Project (www.legato-project.eu), grant agreement no 780681.

\bibliographystyle{ACM-Reference-Format}
\bibliography{icpp}


\begin{thebibliography}{20}


\ifx \showCODEN    \undefined \def \showCODEN     #1{\unskip}     \fi
\ifx \showDOI      \undefined \def \showDOI       #1{#1}\fi
\ifx \showISBNx    \undefined \def \showISBNx     #1{\unskip}     \fi
\ifx \showISBNxiii \undefined \def \showISBNxiii  #1{\unskip}     \fi
\ifx \showISSN     \undefined \def \showISSN      #1{\unskip}     \fi
\ifx \showLCCN     \undefined \def \showLCCN      #1{\unskip}     \fi
\ifx \shownote     \undefined \def \shownote      #1{#1}          \fi
\ifx \showarticletitle \undefined \def \showarticletitle #1{#1}   \fi
\ifx \showURL      \undefined \def \showURL       {\relax}        \fi
\providecommand\bibfield[2]{#2}
\providecommand\bibinfo[2]{#2}
\providecommand\natexlab[1]{#1}
\providecommand\showeprint[2][]{arXiv:#2}

\bibitem[\protect\citeauthoryear{Blelloch, Gibbons, and Matias}{Blelloch
  et~al\mbox{.}}{1999}]%
        {blelloch-jacm99}
\bibfield{author}{\bibinfo{person}{Guy~E. Blelloch},
  \bibinfo{person}{Phillip~B. Gibbons}, {and} \bibinfo{person}{Yossi Matias}.}
  \bibinfo{year}{1999}\natexlab{}.
\newblock \showarticletitle{Provably Efficient Scheduling for Languages with
  Fine-grained Parallelism}.
\newblock \bibinfo{journal}{\emph{J. ACM}} \bibinfo{volume}{46},
  \bibinfo{number}{2} (\bibinfo{date}{March} \bibinfo{year}{1999}),
  \bibinfo{pages}{281--321}.
\newblock
\showISSN{0004-5411}
\urldef\tempurl%
\url{https://doi.org/10.1145/301970.301974}
\showDOI{\tempurl}


\bibitem[\protect\citeauthoryear{Blumofe and Leiserson}{Blumofe and
  Leiserson}{1999}]%
        {blumofe-jacm99}
\bibfield{author}{\bibinfo{person}{Robert~D. Blumofe} {and}
  \bibinfo{person}{Charles~E. Leiserson}.} \bibinfo{year}{1999}\natexlab{}.
\newblock \showarticletitle{Scheduling Multithreaded Computations by Work
  Stealing}.
\newblock \bibinfo{journal}{\emph{J. ACM}} \bibinfo{volume}{46},
  \bibinfo{number}{5} (\bibinfo{date}{Sept.} \bibinfo{year}{1999}),
  \bibinfo{pages}{720--748}.
\newblock


\bibitem[\protect\citeauthoryear{Chakrabarti, Demmel, and Yelick}{Chakrabarti
  et~al\mbox{.}}{1997}]%
        {chakrabarti-jpdc97}
\bibfield{author}{\bibinfo{person}{Soumen Chakrabarti}, \bibinfo{person}{James
  Demmel}, {and} \bibinfo{person}{Katherine Yelick}.}
  \bibinfo{year}{1997}\natexlab{}.
\newblock \showarticletitle{Models and Scheduling Algorithms for Mixed Data and
  Task Parallel Programs}.
\newblock \bibinfo{journal}{\emph{J. Parallel and Distrib. Comput.}}
  \bibinfo{volume}{47}, \bibinfo{number}{2} (\bibinfo{year}{1997}),
  \bibinfo{pages}{168 -- 184}.
\newblock
\showISSN{0743-7315}
\urldef\tempurl%
\url{https://doi.org/10.1006/jpdc.1997.1413}
\showDOI{\tempurl}


\bibitem[\protect\citeauthoryear{Chen, Gibbons, Kozuch, Liaskovitis, Ailamaki,
  Blelloch, Falsafi, Fix, Hardavellas, Mowry, and Wilkerson}{Chen
  et~al\mbox{.}}{2007}]%
        {chen-spaa07}
\bibfield{author}{\bibinfo{person}{Shimin Chen}, \bibinfo{person}{Phillip~B.
  Gibbons}, \bibinfo{person}{Michael Kozuch}, \bibinfo{person}{Vasileios
  Liaskovitis}, \bibinfo{person}{Anastassia Ailamaki}, \bibinfo{person}{Guy~E.
  Blelloch}, \bibinfo{person}{Babak Falsafi}, \bibinfo{person}{Limor Fix},
  \bibinfo{person}{Nikos Hardavellas}, \bibinfo{person}{Todd~C. Mowry}, {and}
  \bibinfo{person}{Chris Wilkerson}.} \bibinfo{year}{2007}\natexlab{}.
\newblock \showarticletitle{{Scheduling Threads for Constructive Cache Sharing
  on CMPs}}.
\newblock \bibinfo{journal}{\emph{Symposium on Parallel Algorithms and
  Architectures}}.
\newblock


\bibitem[\protect\citeauthoryear{Cheng}{Cheng}{2010}]%
        {cheng2010}
\bibfield{author}{\bibinfo{person}{Hui Cheng}.}
  \bibinfo{year}{2010}\natexlab{}.
\newblock \showarticletitle{{A High Efficient Task Scheduling Algorithm Based
  on Heterogeneous Multi-Core Processor}}.
\newblock \bibinfo{journal}{\emph{2010 2nd International Workshop on Database
  Technology and Applications}} \bibinfo{number}{3} (\bibinfo{year}{2010}),
  \bibinfo{pages}{1--4}.
\newblock
\showISBNx{978-1-4244-6975-8}
\urldef\tempurl%
\url{https://doi.org/10.1109/DBTA.2010.5659041}
\showDOI{\tempurl}


\bibitem[\protect\citeauthoryear{Chronaki, Rico, Badia, Ayguad{\'e}, Labarta,
  and Valero}{Chronaki et~al\mbox{.}}{2015}]%
        {chronaki-ics15}
\bibfield{author}{\bibinfo{person}{Kallia Chronaki}, \bibinfo{person}{Alejandro
  Rico}, \bibinfo{person}{Rosa~M. Badia}, \bibinfo{person}{Eduard Ayguad{\'e}},
  \bibinfo{person}{Jes\'{u}s Labarta}, {and} \bibinfo{person}{Mateo Valero}.}
  \bibinfo{year}{2015}\natexlab{}.
\newblock \showarticletitle{Criticality-Aware Dynamic Task Scheduling for
  Heterogeneous Architectures}. In \bibinfo{booktitle}{\emph{Proceedings of the
  29th ACM on International Conference on Supercomputing}}
  \emph{(\bibinfo{series}{ICS '15})}. \bibinfo{publisher}{ACM},
  \bibinfo{address}{New York, NY, USA}, \bibinfo{pages}{329--338}.
\newblock
\showISBNx{978-1-4503-3559-1}
\urldef\tempurl%
\url{https://doi.org/10.1145/2751205.2751235}
\showDOI{\tempurl}


\bibitem[\protect\citeauthoryear{Chronaki, Rico, Casas, Moretó, Badia,
  Ayguadé, Labarta, and Valero}{Chronaki et~al\mbox{.}}{2017}]%
        {chronaki-tpds17}
\bibfield{author}{\bibinfo{person}{K. Chronaki}, \bibinfo{person}{A. Rico},
  \bibinfo{person}{M. Casas}, \bibinfo{person}{M. Moretó},
  \bibinfo{person}{R.~M. Badia}, \bibinfo{person}{E. Ayguadé},
  \bibinfo{person}{J. Labarta}, {and} \bibinfo{person}{M. Valero}.}
  \bibinfo{year}{2017}\natexlab{}.
\newblock \showarticletitle{Task Scheduling Techniques for Asymmetric
  Multi-Core Systems}.
\newblock \bibinfo{journal}{\emph{IEEE Transactions on Parallel and Distributed
  Systems}} \bibinfo{volume}{28}, \bibinfo{number}{7} (\bibinfo{date}{July}
  \bibinfo{year}{2017}), \bibinfo{pages}{2074--2087}.
\newblock
\showISSN{1045-9219}
\urldef\tempurl%
\url{https://doi.org/10.1109/TPDS.2016.2633347}
\showDOI{\tempurl}


\bibitem[\protect\citeauthoryear{Goglin}{Goglin}{2016}]%
        {goglin-hal16}
\bibfield{author}{\bibinfo{person}{Brice Goglin}.}
  \bibinfo{year}{2016}\natexlab{}.
\newblock \bibinfo{booktitle}{\emph{{Towards the Structural Modeling of the
  Topology of next-generation heterogeneous cluster Nodes with hwloc}}}.
\newblock \bibinfo{type}{Research Report}. \bibinfo{institution}{{Inria}}.
\newblock
\urldef\tempurl%
\url{https://hal.inria.fr/hal-01400264}
\showURL{%
\tempurl}


\bibitem[\protect\citeauthoryear{Iyer, Zhao, Guo, Illikkal, Makineni, Newell,
  Solihin, Hsu, and Reinhardt}{Iyer et~al\mbox{.}}{2007}]%
        {iyer-sigmetrics07}
\bibfield{author}{\bibinfo{person}{Ravi Iyer}, \bibinfo{person}{Li Zhao},
  \bibinfo{person}{Fei Guo}, \bibinfo{person}{Ramesh Illikkal},
  \bibinfo{person}{Srihari Makineni}, \bibinfo{person}{Don Newell},
  \bibinfo{person}{Yan Solihin}, \bibinfo{person}{Lisa Hsu}, {and}
  \bibinfo{person}{Steve Reinhardt}.} \bibinfo{year}{2007}\natexlab{}.
\newblock \showarticletitle{QoS Policies and Architecture for Cache/Memory in
  CMP Platforms}. In \bibinfo{booktitle}{\emph{Proceedings of the 2007 ACM
  SIGMETRICS International Conference on Measurement and Modeling of Computer
  Systems}} \emph{(\bibinfo{series}{SIGMETRICS '07})}.
  \bibinfo{publisher}{ACM}, \bibinfo{address}{New York, NY, USA},
  \bibinfo{pages}{25--36}.
\newblock
\showISBNx{978-1-59593-639-4}
\urldef\tempurl%
\url{https://doi.org/10.1145/1254882.1254886}
\showDOI{\tempurl}


\bibitem[\protect\citeauthoryear{Koufaty, Reddy, and Hahn}{Koufaty
  et~al\mbox{.}}{2010}]%
        {Koufaty}
\bibfield{author}{\bibinfo{person}{David Koufaty}, \bibinfo{person}{Dheeraj
  Reddy}, {and} \bibinfo{person}{Scott Hahn}.} \bibinfo{year}{2010}\natexlab{}.
\newblock \showarticletitle{{Bias Scheduling in Heterogeneous Multi-core
  Architectures General Terms Algorithms, Performance}}. In
  \bibinfo{booktitle}{\emph{Proceedings of the 5th European conference on
  Computer systems}}. \bibinfo{pages}{125--138}.
\newblock


\bibitem[\protect\citeauthoryear{Le~Sueur and Heiser}{Le~Sueur and
  Heiser}{2010}]%
        {dvfs}
\bibfield{author}{\bibinfo{person}{Etienne Le~Sueur} {and}
  \bibinfo{person}{Gernot Heiser}.} \bibinfo{year}{2010}\natexlab{}.
\newblock \showarticletitle{Dynamic Voltage and Frequency Scaling: The Laws of
  Diminishing Returns}. In \bibinfo{booktitle}{\emph{Proceedings of the 2010
  International Conference on Power Aware Computing and Systems}}
  \emph{(\bibinfo{series}{HotPower'10})}. \bibinfo{publisher}{USENIX
  Association}, \bibinfo{address}{Berkeley, CA, USA}, \bibinfo{pages}{1--8}.
\newblock
\urldef\tempurl%
\url{http://dl.acm.org/citation.cfm?id=1924920.1924921}
\showURL{%
\tempurl}


\bibitem[\protect\citeauthoryear{Pericas}{Pericas}{2018}]%
        {pericas-taco18}
\bibfield{author}{\bibinfo{person}{Miquel Pericas}.}
  \bibinfo{year}{2018}\natexlab{}.
\newblock \showarticletitle{Elastic Places: an adaptive resource manager for
  scalable and portable performance}.
\newblock \bibinfo{journal}{\emph{ACM Transactions on Architecture and Code
  Optimization}} \bibinfo{volume}{15}, \bibinfo{number}{2}
  (\bibinfo{date}{June} \bibinfo{year}{2018}).
\newblock
\urldef\tempurl%
\url{https://doi.org/10.1145/3185458}
\showDOI{\tempurl}


\bibitem[\protect\citeauthoryear{Qureshi and Patt}{Qureshi and Patt}{2006}]%
        {qureshi-micro06}
\bibfield{author}{\bibinfo{person}{Moinuddin~K. Qureshi} {and}
  \bibinfo{person}{Yale~N. Patt}.} \bibinfo{year}{2006}\natexlab{}.
\newblock \showarticletitle{Utility-Based Cache Partitioning: A Low-Overhead,
  High-Performance, Runtime Mechanism to Partition Shared Caches}. In
  \bibinfo{booktitle}{\emph{Proceedings of the 39th Annual IEEE/ACM
  International Symposium on Microarchitecture}} \emph{(\bibinfo{series}{MICRO
  39})}. \bibinfo{publisher}{IEEE Computer Society},
  \bibinfo{address}{Washington, DC, USA}, \bibinfo{pages}{423--432}.
\newblock
\showISBNx{0-7695-2732-9}
\urldef\tempurl%
\url{https://doi.org/10.1109/MICRO.2006.49}
\showDOI{\tempurl}


\bibitem[\protect\citeauthoryear{Redmon, Divvala, Girshick, and Farhadi}{Redmon
  et~al\mbox{.}}{2016}]%
        {redmon-cvpr16}
\bibfield{author}{\bibinfo{person}{Joseph Redmon}, \bibinfo{person}{Santosh
  Divvala}, \bibinfo{person}{Ross Girshick}, {and} \bibinfo{person}{Ali
  Farhadi}.} \bibinfo{year}{2016}\natexlab{}.
\newblock \showarticletitle{You Only Look Once: Unified, Real-Time Object
  Detection}.
\newblock \bibinfo{journal}{\emph{2016 IEEE Conference on Computer Vision and
  Pattern Recognition (CVPR)}} (\bibinfo{date}{Jun} \bibinfo{year}{2016}).
\newblock
\showISBNx{9781467388511}
\urldef\tempurl%
\url{https://doi.org/10.1109/cvpr.2016.91}
\showDOI{\tempurl}


\bibitem[\protect\citeauthoryear{Sbirlea, Agrawal, and Sarkar}{Sbirlea
  et~al\mbox{.}}{2015}]%
        {sbirlea-europar15}
\bibfield{author}{\bibinfo{person}{Alina Sbirlea}, \bibinfo{person}{Kunal
  Agrawal}, {and} \bibinfo{person}{Vivek Sarkar}.}
  \bibinfo{year}{2015}\natexlab{}.
\newblock \showarticletitle{Elastic Tasks: Unifying Task Parallelism and SPMD
  Parallelism with an Adaptive Runtime}.
\newblock In \bibinfo{booktitle}{\emph{Euro-Par 2015: Parallel Processing}}.
  \bibinfo{series}{Lecture Notes in Computer Science},
  Vol.~\bibinfo{volume}{9233}. \bibinfo{publisher}{Springer},
  \bibinfo{pages}{491--503}.
\newblock
\showISBNx{978-3-662-48095-3}
\urldef\tempurl%
\url{https://doi.org/10.1007/978-3-662-48096-0_38}
\showDOI{\tempurl}


\bibitem[\protect\citeauthoryear{Simonyan and Zisserman}{Simonyan and
  Zisserman}{2014}]%
        {simonyan-arxiv14}
\bibfield{author}{\bibinfo{person}{Karen Simonyan} {and}
  \bibinfo{person}{Andrew Zisserman}.} \bibinfo{year}{2014}\natexlab{}.
\newblock \bibinfo{title}{Very Deep Convolutional Networks for Large-Scale
  Image Recognition}.
\newblock
\newblock
\showeprint[arxiv]{cs.CV/1409.1556}


\bibitem[\protect\citeauthoryear{{Topcuoglu} and and}{{Topcuoglu} and
  and}{2002}]%
        {topcuoglu-tpds02}
\bibfield{author}{\bibinfo{person}{H. {Topcuoglu}} {and}
  \bibinfo{person}{S.~{Hariri} and}.} \bibinfo{year}{2002}\natexlab{}.
\newblock \showarticletitle{Performance-effective and low-complexity task
  scheduling for heterogeneous computing}.
\newblock \bibinfo{journal}{\emph{IEEE Transactions on Parallel and Distributed
  Systems}} \bibinfo{volume}{13}, \bibinfo{number}{3} (\bibinfo{date}{March}
  \bibinfo{year}{2002}), \bibinfo{pages}{260--274}.
\newblock
\showISSN{1045-9219}
\urldef\tempurl%
\url{https://doi.org/10.1109/71.993206}
\showDOI{\tempurl}


\bibitem[\protect\citeauthoryear{{Van Craeynest}, Jaleel, Eeckhout, Narvaez,
  and Emer}{{Van Craeynest} et~al\mbox{.}}{2012}]%
        {VanCraeynest2012}
\bibfield{author}{\bibinfo{person}{Kenzo {Van Craeynest}},
  \bibinfo{person}{Aamer Jaleel}, \bibinfo{person}{Lieven Eeckhout},
  \bibinfo{person}{Paolo Narvaez}, {and} \bibinfo{person}{Joel Emer}.}
  \bibinfo{year}{2012}\natexlab{}.
\newblock \showarticletitle{{Scheduling heterogeneous multi-cores through
  performance impact estimation (PIE)}}. In \bibinfo{booktitle}{\emph{Computer
  Architecture (ISCA), 2012 39th Annual International Symposium}}.
  \bibinfo{pages}{213--224}.
\newblock
\showISBNx{9781467304757}
\showISSN{10636897}
\urldef\tempurl%
\url{https://doi.org/10.1109/ISCA.2012.6237019}
\showDOI{\tempurl}


\bibitem[\protect\citeauthoryear{Wimmer and Tr\"{a}ff}{Wimmer and
  Tr\"{a}ff}{2011}]%
        {wimmer-spaa11}
\bibfield{author}{\bibinfo{person}{Martin Wimmer} {and}
  \bibinfo{person}{Jesper~Larsson Tr\"{a}ff}.} \bibinfo{year}{2011}\natexlab{}.
\newblock \showarticletitle{Work-stealing for Mixed-mode Parallelism by
  Deterministic Team-building}. In \bibinfo{booktitle}{\emph{Proceedings of the
  Twenty-third Annual ACM Symposium on Parallelism in Algorithms and
  Architectures}} \emph{(\bibinfo{series}{SPAA '11})}.
  \bibinfo{pages}{105--116}.
\newblock
\showISBNx{978-1-4503-0743-7}
\urldef\tempurl%
\url{https://doi.org/10.1145/1989493.1989507}
\showDOI{\tempurl}


\bibitem[\protect\citeauthoryear{Zhuravlev, Blagodurov, and Fedorova}{Zhuravlev
  et~al\mbox{.}}{2010}]%
        {zhuravlev-asplos10}
\bibfield{author}{\bibinfo{person}{Sergey Zhuravlev}, \bibinfo{person}{Sergey
  Blagodurov}, {and} \bibinfo{person}{Alexandra Fedorova}.}
  \bibinfo{year}{2010}\natexlab{}.
\newblock \showarticletitle{Addressing Shared Resource Contention in Multicore
  Processors via Scheduling}. In \bibinfo{booktitle}{\emph{Proceedings of the
  Fifteenth Edition of ASPLOS on Architectural Support for Programming
  Languages and Operating Systems}} \emph{(\bibinfo{series}{ASPLOS XV})}.
  \bibinfo{publisher}{ACM}, \bibinfo{address}{New York, NY, USA},
  \bibinfo{pages}{129--142}.
\newblock
\showISBNx{978-1-60558-839-1}
\urldef\tempurl%
\url{https://doi.org/10.1145/1736020.1736036}
\showDOI{\tempurl}


\end{thebibliography}

%

\end{document}